\documentstyle[prc,aps,manuscript]{revtex}

\begin{document}

\draft

\title{CALORIC CURVE FOR FINITE NUCLEI IN THOMAS-FERMI THEORY}

\author{J. N. De$^{1*)}$, S. Das Gupta$^{2)}$, 
        S. Shlomo$^{1)}$ and S. K. Samaddar$^{3)}$}

\address{$^{1)}$ Cyclotron Institute, Texas A\&M University,
College Station, TX 77843-3366, USA}
\address{$^{2)}$ Department of Physics, McGill University, 3600
University St., Montreal, PQ, H3A 2T8 Canada}
\address{$^{3)}$ Saha Institute of Nuclear Physics, 1/AF,
Bidhannagar, Calcutta - 700064, India}

\maketitle

\begin{abstract}
In a finite temperature Thomas-Fermi theory with realistic nuclear 
interactions, we construct caloric curves for finite nuclei enclosed 
in a sphere of about $4 - 8$ times the normal nuclear volume.  The 
specific heat capacity $C_v$ shows a peaked structure that is possibly 
indicative of a liquid-gas phase transition in finite nuclear systems.  
\end{abstract}

\vskip 1 cm

\vskip 1 cm


\newpage

The equation of state (E.O.S.) of nuclear matter with realistic effective
interactions shows a typical Van der Waals type behavior and a critical 
temperature of $\approx 15 - 20$ MeV \cite{kwh,jmz,bssd}. 
Supported by the experimental 
observation of a power law behavior in the mass or charge distribution 
in proton \cite{fi,gkj}
and heavy ion induced reactions\cite{chit,lynen}, 
the idea of liquid-gas 
phase transition in nuclear matter or finite nuclear systems 
\cite{jmz,siem,pan,gkm} has gotten 
considerable interest in the literature. Theoretical speculations and 
possible experimental indications of a limiting temperature 
\cite{lb,blv,drss,gue,nat} in finite 
nuclei at $\approx 5 - 7$ MeV, above which the nucleus becomes unstable and 
breaks up into many fragments, also calls for a possible connection 
between the limiting temperature and the phase transition.  Phase 
transitions are normally signalled by peaks in the specific heat at 
constant volume, $C_v$ as temperature increases.  Fragmentation 
calculations in the microcanonical algorithm of Gross \cite{gross}
and in the 
Copenhagen canonical description \cite{bdms,bbims}
show such peaks. Recent calculations 
by Das Gupta et. al. \cite{gupta}
in the lattice gas model for fragmentation also 
show such a structure.  Renewed interest in this subject was further 
fueled by the recent experimental observation \cite{poch}
in the caloric curve 
of a near constancy of temperature in the excitation energy range 
of $\approx 4 - 10$ MeV/nucleon in Au + Au collisions. This prompted us 
to find out whether the trends in the caloric curve as seen in the 
experiment or in fragmentation calculations are reproduced in a finite 
temperature Thomas-Fermi (TF) theory.  To our knowledge this is the 
first calculation of its kind with a realistic effective interaction.
In the context of an exactly solvable Fermion model, Rossignoli et al 
\cite{rpm}
have earlier calculated the specific heat of a finite nucleus in 
the grand canonical mean field theory with Lipkin's model hamiltonian, 
but found no structure in it as a function of temperature.  The 
structure appeared in the canonical calculation, with inclusion of 
correlations.

   In our refined Thomas-Fermi (TF) model, the interaction density is 
calculated with a Seyler-Blanchard type \cite{sb}
momentum and density dependent
finite range two-body effective interaction \cite{drss}. 
The interaction is given by

$$
v_{{\rm eff}}(r,p,\rho)=C_{l,u}[v_{1}(r,p)+v_{2}(r,\rho)]
\eqno (1) 
$$
$$
v_{1}=-(1-p^{2}/b^{2})f({\bf r_{1}},{\bf r_{2}})
$$
$$
v_{2}=d^{2}[\rho_{1}(r_{1})+\rho_{2}(r_{2})]^{n}
f({\bf r_{1}},{\bf r_{2}})
\eqno (2) 
$$
with
$$
f({\bf r_{1}},{\bf r_{2}})=\frac{e^{- \mid {\bf r_{1}}-{\bf r_{2}} \mid } /
a}{ \mid {\bf r_{1}}-{\bf r_{2}} \mid /a} .
\eqno (3) 
$$
Here $a$ is the spatial range and $b$ the strength of repulsion in the 
momentum dependence of the interaction, $r=\mid {\bf r_{1}}-{\bf r_{2}}
\mid $ 
and $p= \mid {\bf p_{1}}-{\bf p_{2}} \mid $ are the relative distance and 
relative momenta of the two interacting nucleons. The subscripts 
$l$ and $u$ in the strength $C$ refer to like pair (n-n or p-p) or
unlike pair (n-p) interaction respectively, $d$ and $n$ are measures
of the strength of the density dependence of the interaction and 
$\rho_1$ and $\rho_2$ are the densities at the sites of the two 
nucleons.

The potential parameters are determined for a fixed value of $n$ from 
a fit of the well-established bulk nuclear properties and the value 
of $n$ is determined \cite{drss} from a fit of the Giant Monopole 
Resonance energies over a broad mass spectrum.  

The Coulomb interaction energy density is given by the sum of the 
direct and exchange terms. They are given by

$$
 \varepsilon_{D}(r)=e^{2} \pi \rho_{p}(r) \:
\int \: dr^{\prime} \, r^{\prime \, 2} \rho_{p}(r^{\prime})
{\rm g}(r,r^{\prime}),
\eqno (4) 
$$
and
$$
 \varepsilon_{{\rm ex}}(r)=-\frac{3e^{2}}{4\pi}(3\pi^{2})^{1/3}
\rho_{p}^{4/3}(r) .
\eqno (5) 
$$
Here $\rho_p(r)$ is the proton density and
$$
 {\rm g}(r,r^{\prime})=\frac{(r+r^{\prime})-\mid r-r^{\prime} \mid}
{rr^{\prime}} .
\eqno (6) 
$$
With the potential chosen, the total energy density at a temperature 
$T$ is then written as
$$
\varepsilon(r)=\sum_{\tau} \, \rho_{\tau}(r) [T \, J_{3/2}(\eta_{\tau}(r))/
J_{1/2}(\eta_{\tau}(r)) {(1- m_{\tau}^{\ast}(r) V_{\tau}^{1}(r))}
+\frac{1}{2}V_{\tau}^{0}(r)]
\eqno (7) 
$$
Here $\tau$ refers to neutron or proton, the $J$'s are the usual 
Fermi integrals, $V_{\tau}^{0}$ is the single particle potential 
( for protons, it includes the Coulomb term), $V_{\tau}^{1}$ is 
the potential term that comes with momentum dependence and is 
associated with the effective mass $m_{\tau}^{\ast}$. The fugacity 
$\eta_{\tau}(r)$ is defined as 
$$
 \eta_{\tau}(r)=[\mu_{\tau}-V_{\tau}^{0}(r)-V_{\tau}^{2}(r)]/T
\eqno (8) 
$$
where $\mu_{\tau}$ is the chemical potential and $V_{\tau}^{2}$ is 
the rearrangement potential that appears for a density-dependent 
interaction. The total energy per particle at any temperature is 
then given by  
$$
  E(T)= \int \, \varepsilon(r) \, {\rm d}^{3} r /A .
\eqno (9) 
$$
Once the interaction energy density is known, the nuclear density can
be obtained self-consistently and other observables of physical interest 
calculated. For details on the finite temperature TF theory, we refer 
to Ref. \cite{drss}. 

Since the continuum states of a nucleus at nonzero temperature are
occupied with a finite probability given by a Fermi factor \cite{balian},
the particle density does not vanish at large distances. The 
observables then depend on the size of the box in which the 
calculations are performed. Guided by the practice that many calculations
for heavy ion collisions are done by imposing that thermalisation occurs 
in a freeze-out volume, we fix a volume and find out the excitation 
energy as a function of temperature which allows for the determination 
of the specific heat at constant volume.

We choose two systems, namely $^{150}$Sm and $^{85}$Kr. In the context
of very heavy ion collisions at intermediate or higher energies, this 
mass range is of experimental interest. The calculations have 
been done for two confinement volumes, one at $V =4.0V_0$ and the other at
$V = 8.0V_0$, where $V_0$ is the normal volume of the nucleus at zero
temperature. The calculations at zero temperature are independent of the
volumes taken; at low temperature of $\approx 1 - 2$ MeV, the observables 
are nearly independent of the volume. As the temperature increases, the 
central density is depleted. In Figure 1, the proton densities for
$^{150}$Sm calculated in the volume $V = 8.0V_0$ are displayed for
four temperatures, $T = 5$ MeV (dashed curve), 
$T = 9$ MeV (dotted curve), $T = 9.5$MeV (dash-dotted curve) and 
$T = 10$ MeV (full curve). 
At $T = 5$ MeV, the central density is depleted by $\approx 4\%$ 
compared to zero temperature density, but has a long thin tail spread 
to the boundary. The behaviours at $T = 9$ and 9.5 MeV are qualitatively 
the same, but with further depletion in the central density and a thicker 
tail. Beyond $T = 9.5$MeV, the change in the density starts being abrupt 
and the whole system looks like a uniform
distribution of matter inside the volume. This is shown by a representative 
density distribution at $T = 10$MeV.  
 The slight bump seen in the outer edge 
of the density is due to the Coulomb force. In Figure 2, the proton 
density at $T = 10$ MeV for the system at $V = 8.0V_0$ (dashed curve) is 
compared with that calculated at $V = 4.0V_0$ (full curve). The density
calculated in smaller volume still shows a structure and the central 
density is depleted by only about $20\%$ even at this high temperature. 

The excitation energy per particle $E^{*}$ is defined as 
$E^{*} = E(T) - E(T=0)$. In Figure 3, we display the 
caloric curve for the system $^{150}$Sm.
The upper dashed curve corresponds to $V = 4.0V_0$ while the lower 
full curve corresponds to $V = 8.0V_0$. At lower density, the excitation 
energy rises faster. For both volumes, initially the temperature rises 
faster with excitation energy, then its rise is slower. For the lower 
density, a kink is observed in the caloric curve at $T \approx  10$ MeV, 
after which the excitation energy rises almost linearly with temperature. 
For the higher density, the kink is much smaller and appears at a somewhat 
higher temperature. In Figure 4, the corresponding specific heats $C_v$ 
defined 
as
$$
 C_{v}=\left( {\rm d}E^{*}/{\rm d} T \right)_{v}
\eqno (10) 
$$
are displayed. Since we use units of MeV for both energy and temperature,
the calculated $C_v$ is dimensionless.
For both volumes, the specific heat shows a peak, the peak being much 
sharper for the case of a larger volume. For the smaller volume, the peak 
is at $T \approx 10.5$ MeV while for the large volume the peak is shifted
down by $\approx 1$ MeV. We believe that the kink in the
caloric curve or the peak in the specific heat are related to a phase
transition in finite nuclei. From our calculations, we find that this 
transition temperature is weakly dependent on the confinement volume 
beyond $V = 8V_0$, e.g., for V as high as $20V_0$, the transition 
temperature is shifted down further by only $\approx 1$MeV. The classical 
value of $C_v = 3/2$ is reached
at $T \approx 11$ MeV for the case with $V = 8.0V_0$ while for the smaller
volume, it is reached at $T \approx 13$ MeV. This is expected as the 
interaction becomes weaker either with increased volume or with 
increased temperature.

In Figure 5, the caloric curve for the lower mass system $^{85}$Kr is
shown. The trends are nearly the same as in Figure 3. Figure 6 displays
the specific heat for this system. In the calculation with  
$V = 4V_0$, a broad bump in the specific heat at $T \approx 11$ MeV 
is seen.  
In calculations with expanded volume ($8V_0$), the system shows a sharp 
peak at 
$T \approx  10.5$ MeV. This peak is, however, not as sharp as the one for 
the heavier system. In calculations on limiting temperature in the model 
of liquid-gas phase equilibrium, the influence of Coulomb forces has 
often been emphasized \cite{bssd,jaq} in the instability of the system.
In the present calculation, we see a relatively small effect on the 
transition temperature. With the Coulomb force switched off, the     
transition temperature is shifted up by$\approx 1$ MeV for both the 
confinement volumes $4V_0$ and $8V_0$ and the matter density becomes
more uniform. This transition temperature is somewhat lower compared to
the critical temperature for asymmetric nuclear matter \cite{bssd} with
isospin asymmetry equal to that of the nucleus. 
  
To summarize, we have calculated the caloric curve and the specific
heat for two systems in a self-consistent Thomas-Fermi theory at two
volumes, namely at 4 and 8 times the normal nuclear volume. The specific 
heat $C_v$ shows a peaked structure possibly signalling a liquid-gas 
phase transition at a temperature of $\approx 10$ MeV  
which is lower than the calculated critical
temperature for infinite nuclear matter but larger compared to the 
calculated limiting temperature for finite real nuclei \cite{drss}.
In simplistic model calculations
\cite{rpm}, it has been shown that the inclusion of correlations brings 
in features reminiscent of a phase transition in a system when no
phase transition is evident in the usual mean field calculation;
it would therefore be interesting to see whether fluctuations with 
two-body correlations bring down the phase transition temperature
obtained in our TF calculation.

\bigskip

The authors acknowledge fruitful discussions with Dr. E. Ramakrishnan.
One of the authors (J. N. D.) gratefully acknowledges the hospitality 
of the Cyclotron Institute, Texas A\&M University where this work was 
completed. This work is supported by the U.S. Department of Energy
under grant DE-FE05-86ER40256, by the Natural Sciences and Engineering
Research Council of Canada and by the U.S. National Science Foundation
under grant PHY-9413872.

\bigskip

* On leave of absence from the Variable Energy Cyclotron Centre,
1/AF, Bidhannagar, Calcutta- 700 064, India

\newpage

\begin{center}
{\bf FIGURE CAPTIONS}
\end{center}
\bigskip

{\bf Fig. 1}
The proton density profile for the system $^{150}$Sm calculated at
four temperatures in the volume $V = 8.0V_0$. The dashed, dotted, 
dash-dot and full lines correspond 
to temperatures $T = 5,9,9.5$ and 10 MeV respectively. 
 
{\bf Fig. 2} 
The proton density profile for the system $^{150}$Sm calculated at
temperature $T = 10$ MeV in two different volumes. The full and dashed 
lines correspond to calculations at $V = 4.0V_0$ and $V = 8.0V_0$, 
respectively.

{\bf Fig. 3} 
The temperature plotted as a function of excitation energy per 
particle (caloric curve) for the system $^{150}$Sm. The dashed curve 
corresponds to calculations with volume $V = 4.0V_0$ while the full 
curve corresponds to $V = 8.0V_0$.  

{\bf Fig. 4}
The specific heat per particle plotted as a function of temperature 
for the system $^{150}$Sm. The dashed curve corresponds to calculations 
with volume $V = 4.0V_0$ while the full curve corresponds to $V = 8.0V_0$.

{\bf Fig. 5}
Same as Figure 3 for the system $^{85}$Kr.

{\bf Fig. 6}
Same as Figure 4 for the system $^{85}$Kr.

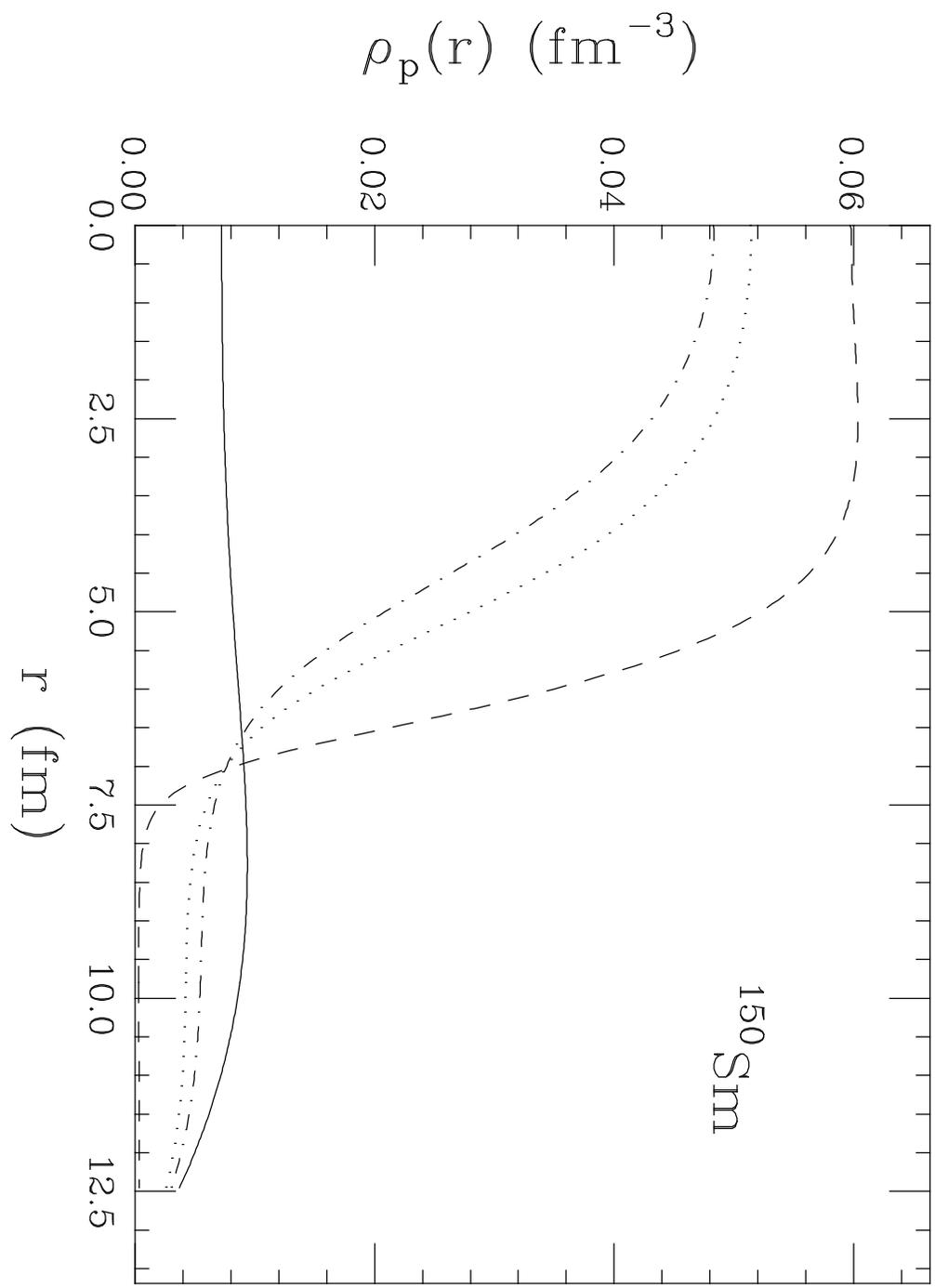
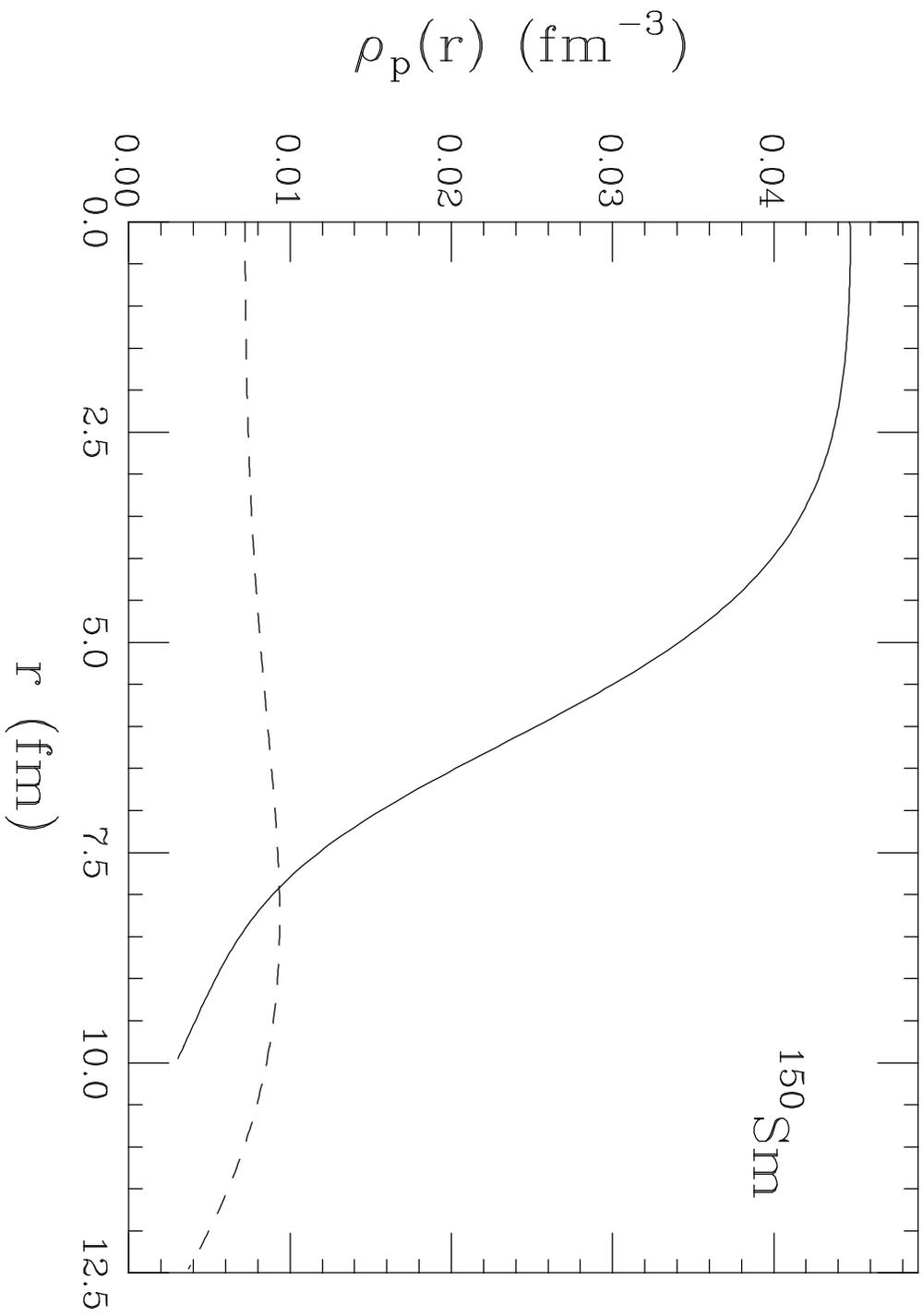
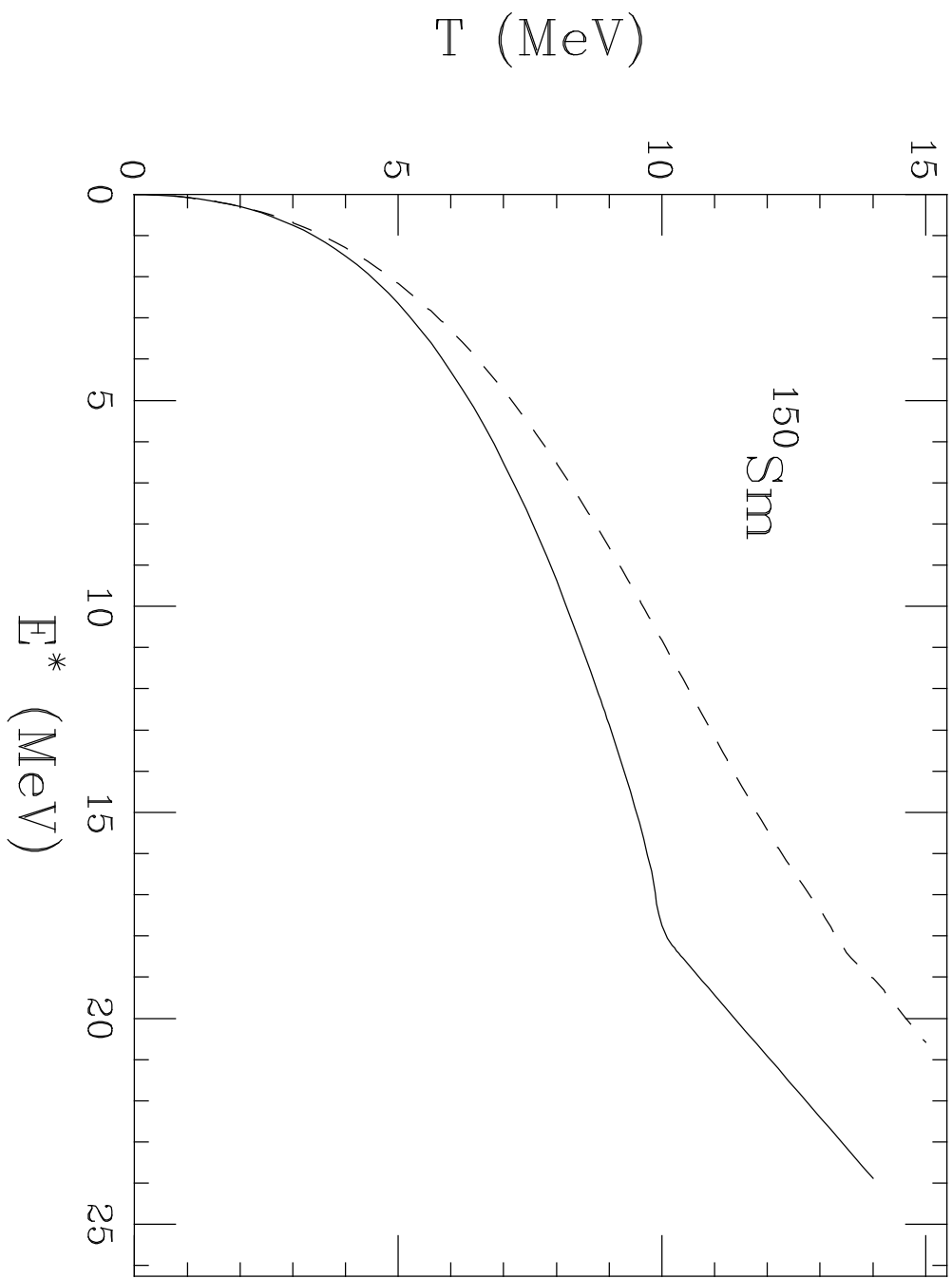
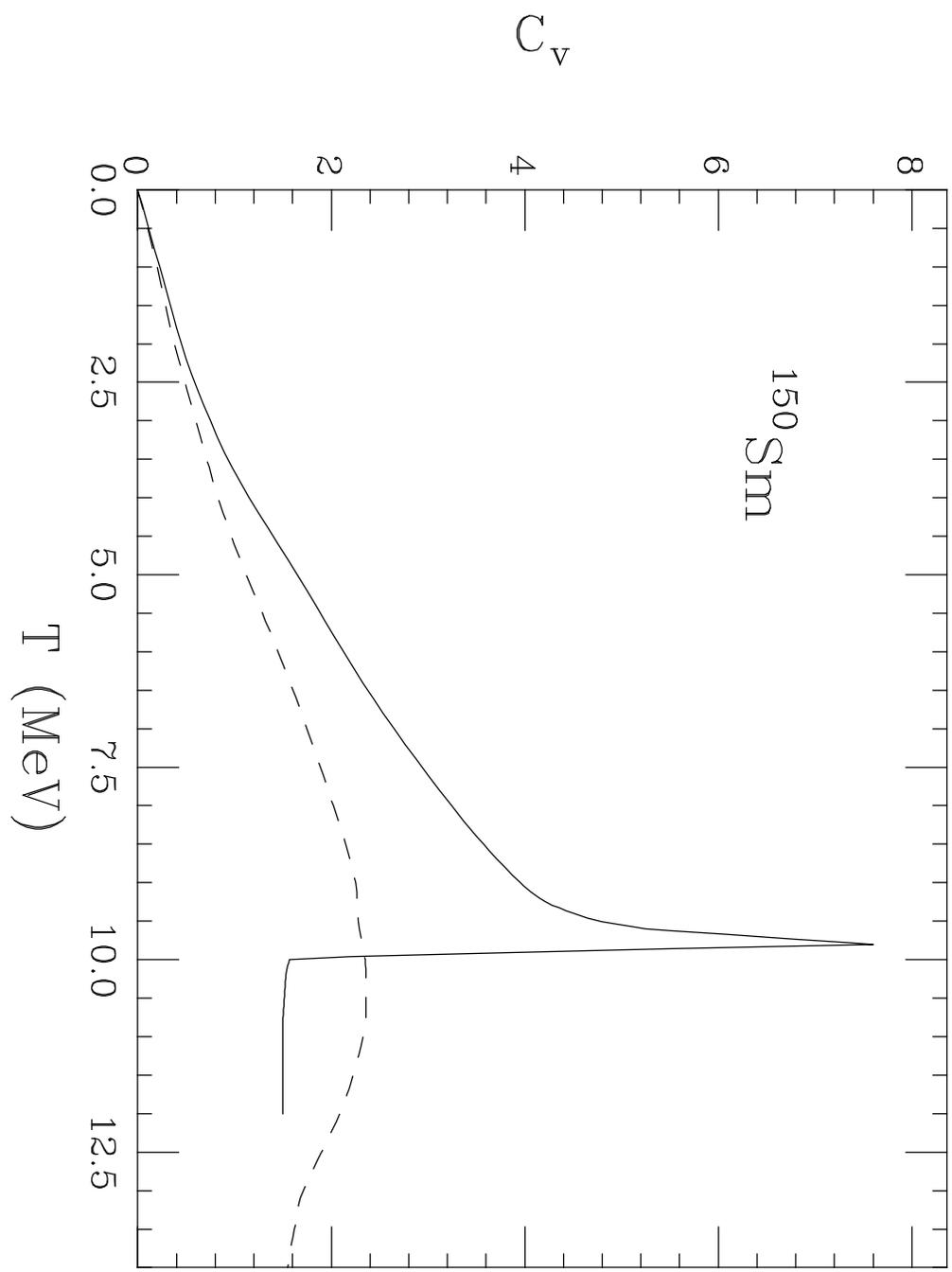
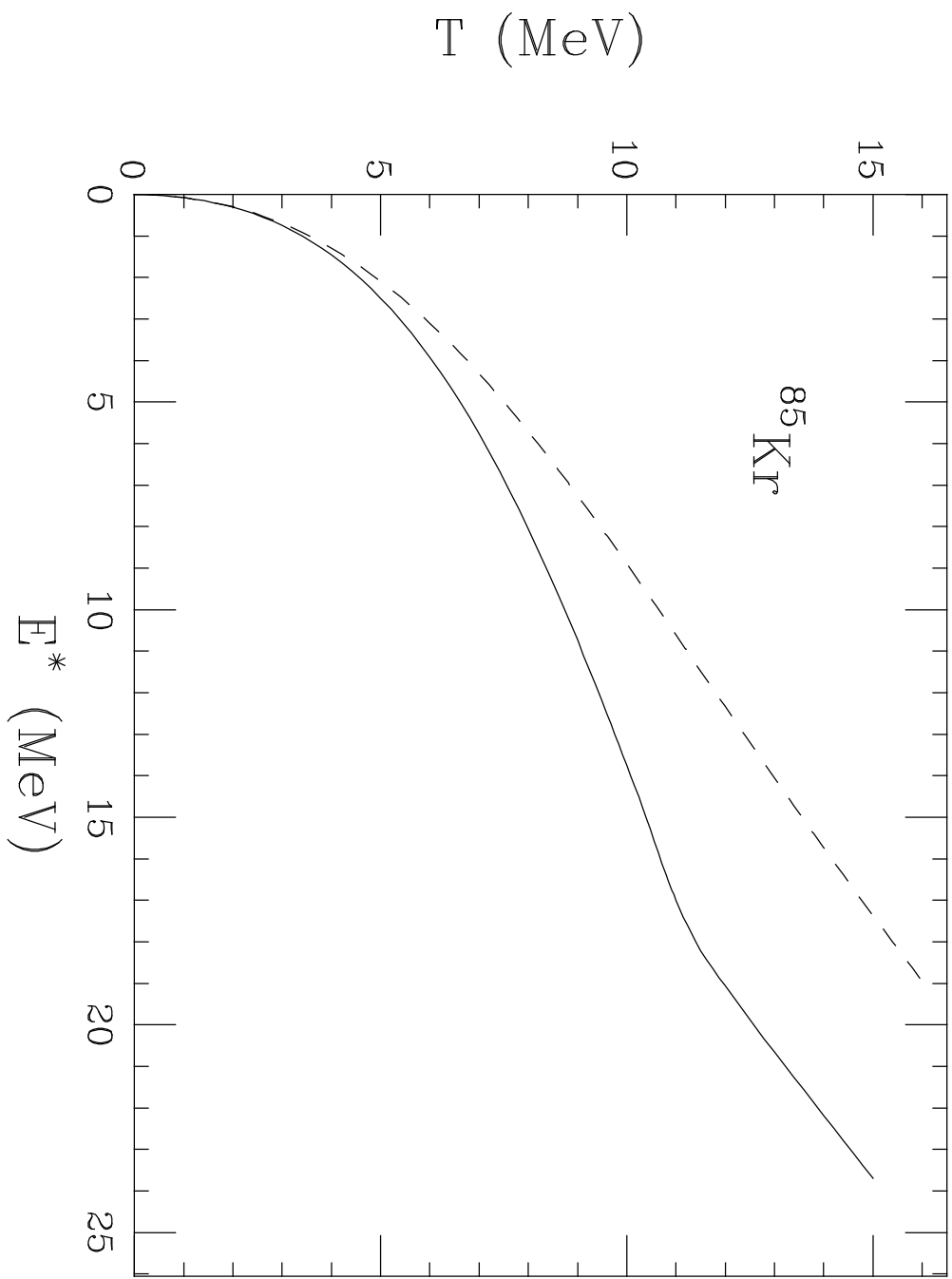
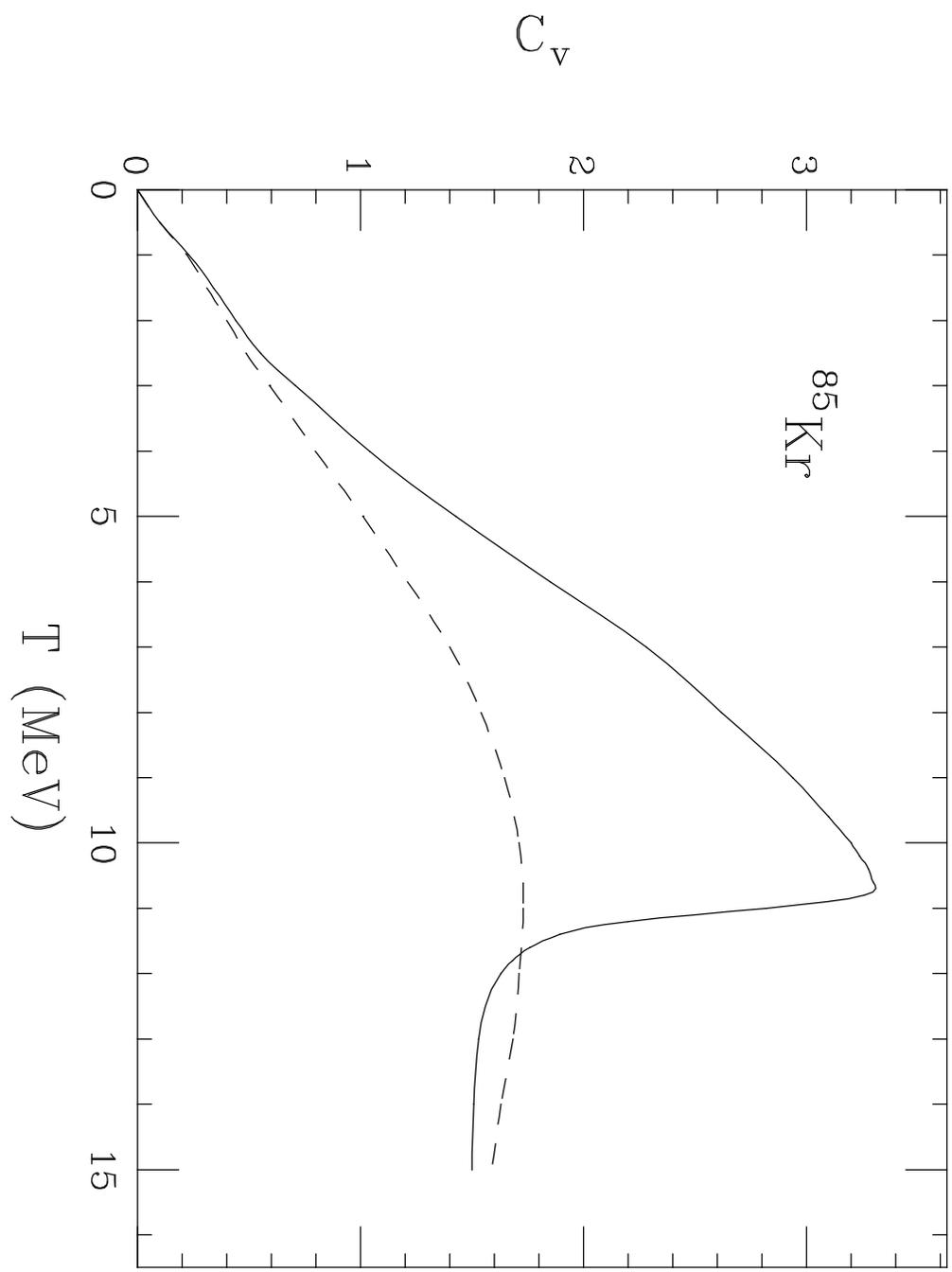
\end{document}